\documentclass[aps,prl,reprint,twocolumn]{revtex4}
\usepackage{blindtext}
\usepackage{epsfig}
\usepackage{color}
\usepackage{amsmath}
\usepackage{amsfonts}
\usepackage{amssymb}
\usepackage{graphicx}%
\usepackage{url}
\usepackage[breaklinks]{hyperref}
\usepackage[english]{babel}
\usepackage{esvect}
\usepackage[normalem]{ulem}
\usepackage[inline]{enumitem}
\usepackage{xfrac}
\usepackage{dsfont}
\usepackage{bm}
\usepackage{titlesec}

\titleformat{\section}[runin]{\normalfont \itshape}{}{}{}[\hspace{8pt}{--}]

\hypersetup{colorlinks,citecolor=blue,filecolor=blue,linkcolor=blue,urlcolor=blue}

\newcommand{\ket}[1]{\vert #1 \rangle}

\begin{document}

\title{Topological Zak Phase in Strongly-Coupled LC Circuits}

\author{Tal Goren$^1$, Kirill Plekhanov$^{1,2}$, F\'{e}licien Appas$^1$, Karyn Le Hur$^1$}
\affiliation{$^1$Centre de Physique Th\'{e}orique, \'{E}cole
  Polytechnique,CNRS, Universit\'{e} Paris-Saclay,  91128 Palaiseau
  C\'{e}dex, France}
\affiliation{$^2$LPTMS, CNRS, Univ. Paris-Sud, Universit\'e
  Paris-Saclay, 91405 Orsay, France}

\begin{abstract}
  We show the emergence of topological Bogoliubov bosonic
  excitations in the relatively strong coupling limit of an LC
  (inductance-capacitance) one-dimensional quantum circuit. This
  dimerized chain model reveals a ${\mathbb Z}_2$ local
  symmetry as a result of the counter-rotating wave (pairing)
  terms. The topology is protected by the sub-lattice symmetry,
  represented by an anti-unitary transformation.  We present a method to    measure the winding of the topological Zak phase across the Brillouin zone by a reflection measurement of (microwave) light. Our method probes bulk quantities and can be implemented even in small systems. We study the robustness of edge modes towards disorder. 
  \end{abstract}

\date{\today}
\maketitle

Topological Bloch bands are characterized by a topological number which is manifested in the appearance of protected edge states. For non-interacting fermions this results in
the celebrated integer quantum Hall effect~\cite{thouless1982}, conducting surface states of topological
insulators~\cite{Hasan2010, bernevig2013} and
semimetals~\cite{Hasan2010, Burkov2016}. Topological properties can also be
accessed with bosonic systems such as cold
atoms~\cite{atala2013}, photons~\cite{lu2014, Review_2016, Jens} and
polaritons~\cite{Bloch2017_pol}.

The Su-Schrieffer-Heeger (SSH) model defined on the dimerized one-dimensional lattice with two sites per unit cell is one of the
simplest models demonstrating topological properties~\cite{SSH1979,asboth2016}. The edge states are protected by the bipartite nature of the system (particles can  hop only between the two sublattices). 
Arbitrary long range interaction which respect the bipartite nature of
the SSH model may change the topological number but not the robustness
of the edge states to disorder~\cite{chen2017}. The topology of the SSH model is described by the Zak phase \cite{Zak1989}, which was measured in cold
atoms by introducing an artificial gauge field and mimicking Bloch
oscillations~\cite{atala2013}, in photonic quantum walk~\cite{cardano2017} and in photonic crystals \cite{Wang2016}. 
The midgap edge stated were observed in dielectric resonators \cite{Bellec}, polariton systems \cite{Jbloch2017} and classical LC chains \cite{Wurzburg}, and their wave function was explored in Ref.~\cite{meier2016,Wurzburg}. 
A two-leg ladder of SSH chains supports fractional excitations and shows a rich topology with two types of corner edge states~\cite{Zhang2017,
  Li2017}. The ladder is a one-dimensional version of a two-dimensional quadrupole insulator~\cite{Bernevig2017} realized
in Refs.~\cite{Huber2017, Imhof2017, Peterson2017}. In the non-linear
regime a dimerized chain shows topologically enforced
bifurcations~\cite{Engelhardt2017}.

Here, we study topology in the strong coupling regime of quantum circuits in which the rotating wave approximation (RWA) is not applicable, leading to the appearance of counter rotating (pairing) terms. Such strong coupling limit also leads to the evolution of the Jaynes-Cummings model towards the Rabi model when describing the qubit-cavity interaction~\cite{Review_2016,niemczyk2010}, and to the super radiant phase in
Dicke model with a macroscopic number of photons in the ground state~\cite{eckert2007dicke, marco} but has not been studied in the framework of topological systems. 
We start by showing that the counter rotating terms do not change the topology of the system although they modify the nature of the excitations from pure particles to Bogoliubov quasi-particles. We also find the wave-function of the midgap states demonstrating their edge localization. Then we propose a method to observe the winding of the toplogical Zak phase and probe the strong coupling regime via a light scattering
experiment. Topological Bogoliubov excitations in other interacting
bosonic systems have been studied in Refs.~\cite{Eng2015,
  Engelhardt2017, Alex, Refael2016}.

\begin{figure}
  \begin{center}
    \includegraphics[width=0.49\textwidth]{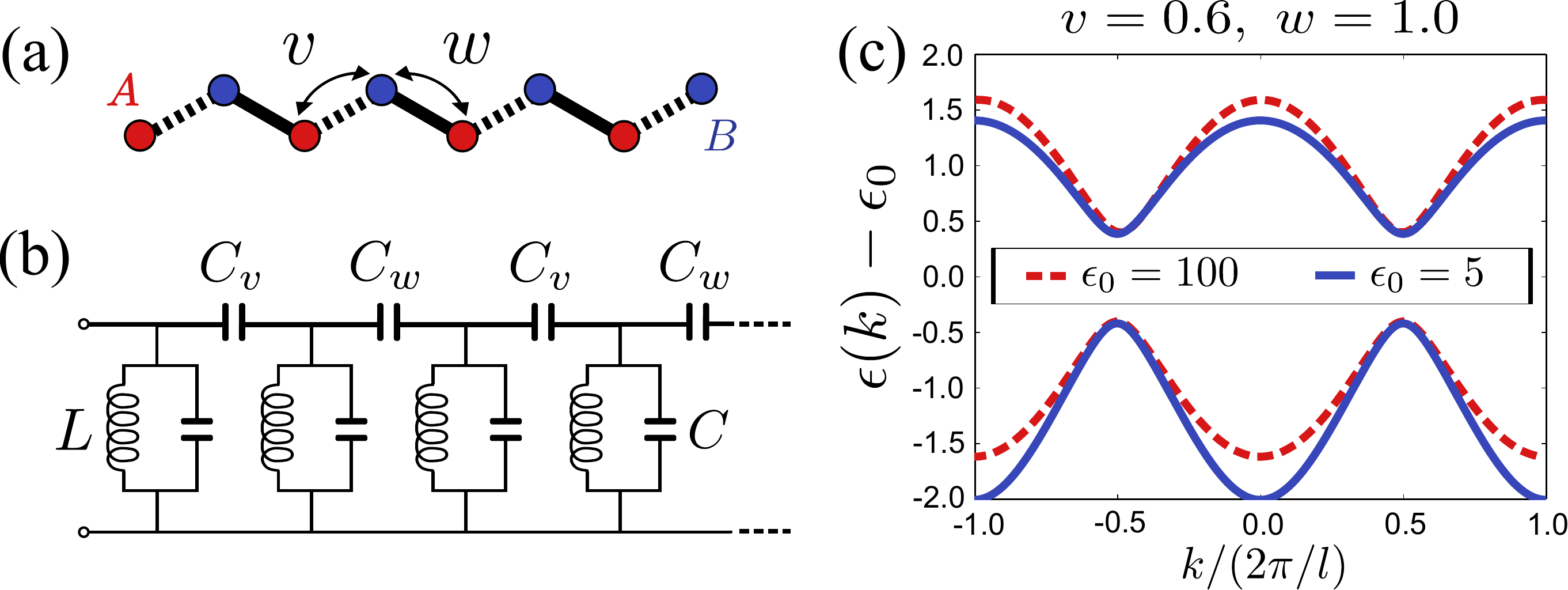}
  \end{center}
  \vskip -0.5cm \protect\caption[band structure]
  {(color online) \it a) Schematic of the dimerized lattice  with two sublattices $A$ and $B$ and alternating couplings $(v,w)$. b) Dimer chain of LC oscillators described by the Hamiltonian
    \eqref{H_main}. c) Bulk band structure in the strong (solid blue) and weak (dashed red) coupling regimes.}
    \label{band_struct}
    \vskip -0.5cm
\end{figure}

In the following, we study a chain of $LC$ resonators with alternating capacitive coupling (see Fig.~\ref{band_struct}). The unit cell of the chain is composed of two sites, denoted by indices $a$ and
$b$ that belong respectively to sublattices $A$ and $B$. Similar systems have been achieved experimentally~\cite{Jia} and some topological aspects can already be measured in the classical limit (without quantification of the phase and charge in the circuit)
\cite{Huber2017, Imhof2017, Peterson2017, Glazman2015}.

The charge on the capacitor of a quantum LC resonator can be quantized introducing bosonic creation and annihilation operators $Q\propto (a+a^\dagger)$ \cite{devoret1997}. The capacitive coupling between resonators $i$ and $i+1$ is $\propto {Q}_i {Q}_{i+1}$. In this description the dimer chain of $LC$ resonators leads to the following Hamiltonian
\begin{align}
  \label{H_main}
  H &= \sum_n{\epsilon_0\left( a_n^\dagger
      a_n +b_n^\dagger b_n \right)}
      +\sum_n{v\left(a_n +a_n^\dagger \right)\left(b_n +b_n^\dagger
      \right)}
  \nonumber \\
  & +\sum_n{w\left(a_{n+1} +a_{n+1}^\dagger \right)\left(b_n
      +b_n^\dagger \right)},
\end{align}
where $(a_n^\dagger,b_n^\dagger)$ are the creation operators in the $n$th unit cell,
$\epsilon_0=\sqrt{\frac{1}{LC}}\sqrt{\frac{C+C_v+C_w}{C}}$ is the
frequency of the resonators and
$v,w=\epsilon_0\sqrt{\frac{C_{v,w}}{C+C_v+C_w}}$ are the hopping
matrix elements.  The inductance $L$ and capacitances $C$, $C_v$ and
$C_w$ are defined in Fig.~\ref{band_struct}. In the interaction
picture, the pairing terms $a^\dagger b^\dagger$, $a b$ oscillate at
twice the resonator frequency $2\epsilon_0$.  When this frequency is
much larger than the hopping amplitudes $\epsilon_0\gg (v,w)$ (i.e. weak
coupling), the RWA neglects these fast rotating terms.  In the strong
coupling regime, where the hopping amplitudes are of the same order of
magnitude as the resonator's frequency $\epsilon_0\sim (v,w)$, the RWA
breaks down thereupon the contribution of the pairing terms
$a^\dagger b^\dagger$, $a b$ must be incorporated. As a result, the
excitations of the system are Bogoliubov quasi-particles, i.e. a
combination of a particle ($a,b$) and anti-particle ($a^\dagger,b^\dagger$). 
Including the pairing terms reduces the $U(1)$ symmetry of the SSH model to a ${\mathbb Z}_2$ pairing symmetry characterized by the transformation $a_n\rightarrow -a_n$ and $b_n\rightarrow -b_n$ in the Hamiltonian,
similarly to the Dicke model. It is important to note that in the
present circuit a super-radiant quantum phase transition cannot occur
since formally $(v,w)<\epsilon_0$.

For periodic boundary conditions the Hamiltonian can be rewritten in
the momentum space as
\begin{align}
  H &= \sum_{k>0}\Phi_{k}^{\dagger}H\left(k\right)\Phi_{k}
  \nonumber \\
  H(k) &= \epsilon_0\mathbb{I}_4 +M(k)\otimes
         \begin{pmatrix}
           1 & 1\\
           1 & 1
         \end{pmatrix}
  \\
  M(k) &=
         \begin{pmatrix}
           0 & h\left(k\right)e^{-i\varphi(k)}\\
           h\left(k\right)e^{i\varphi(k)} & 0
         \end{pmatrix}, \nonumber            
\end{align}
where we have defined
$\Phi_{k}^{\dagger}=\left(a_{k}^{\dagger},b_{k}^{\dagger},a_{-k},b_{-k}\right)$
and $h(k)e^{i\varphi(k)}\equiv v+we^{-ikl}$, $k$ being the lattice
momenta and $l$ is the length of a unit cell; $\mathbb{I}_n$ is the
$n \times n$ identity matrix.  The system is invariant under time
reversal $H(k)=H(-k)^*$, and inversion
$\sigma_1\otimes \mathbb{I}_2 H(-k)\sigma_1\otimes\mathbb{I}_2=H(k)$
symmetries. The sublattice symmetry is the product of these two
symmetries, represented by the anti-unitary transformation
$A=\sigma_1\otimes\mathbb{I}_2\mathcal{K}$, where $\sigma_1$ is the
first Pauli matrix and $\mathcal{K}$ is the complex conjugate operator
$\mathcal{K}z=z^*\mathcal{K}$. The invariance to the sublattice
symmetry $AH(k)=H(k)A$ ensures that the diagonal terms in $M(k)$ will
vanish.

The Hamiltonian can be diagonalized by a Bogoliubov transformation
\begin{equation}
  H = \sum_k{\left[\epsilon_+(k)\gamma_k^{+\dagger}\gamma_k^+
      +\epsilon_-(k)\gamma_k^{-\dagger}\gamma_k^-\right]},
\end{equation}
where
\begin{equation}
  \epsilon_{\pm}(k) = \sqrt{\epsilon_0^2 \pm 2\epsilon_0 h(k)}.
\end{equation}
Due to inversion
symmetry the spectrum is symmetric under $k\rightarrow -k$. The
Bogoliubov quasi-particles operators are given by
\begin{align} \label{eq:basis}
  \gamma_k^{\pm} &= \bm{\psi_k^{\pm}}\cdot(a_k,a_{-k}^\dagger
                   ,b_k,b_{-k}^\dagger) \nonumber
  \\
  \bm{\psi_k^{\pm}} &= \frac{1}{\sqrt{2}}\left(\pm
                     e^{i\varphi(k)}\bold{v_k^\pm},\bold{v_k^\pm}\right)
  \\
  \bold{v_k^\pm}&=(\cosh\eta_k^\pm,\sinh\eta_k^\pm), \nonumber
\end{align}
where the Bogoliubov rotation $\eta_k^\pm$ is defined by
$\tanh2\eta_k^\pm =\pm h(k) / \epsilon_0 \pm h(k)$.  Indeed in the
weak coupling limit $h(k)\ll\epsilon_0$, the spectrum of the SSH chain
is reproduced: $\epsilon_\pm(k) \rightarrow \epsilon_0 \pm h(k)$ and the
excitations are particle-like $\bold{v_k^\pm}\rightarrow (1,0)$.  The band
structure is plotted in Fig.~\ref{band_struct} in the strong and weak
coupling regimes.

Unlike the SSH model, the chiral symmetry ($U=\sigma_3\otimes \mathbb{I}_2$) is explicitly broken by the on-site energy term $\epsilon_0$ ($UH(k)U\neq -H(k)$), therefore the spectrum is not symmetric with respect to $\epsilon_0$. However, in the weak coupling
limit the spectrum is symmetric since the transformation that diagonalizes the Hamiltonian is unitary. In
contrast, in the strong coupling regime, the symmetry is broken due to
the symplecticity of the Bogoliubov transformation (see Fig. ~\ref{band_struct}).

Detecting whether the system is in the strong or the weak coupling
regime can therefore be achieved by studying  the asymmetry of the two bands of the energy
spectrum. The difference between the two bandwidths (see Fig. (2d)), to lowest order
in $v$ and $w$, is
\begin{equation}
  \label{eq:bandwidths}
  \mathcal{W}^- - \mathcal{W}^+ = \frac{4wv}{\epsilon_{0}},
\end{equation}
which vanishes in the weak coupling limit.

In the strong coupling regime, the ground state of the system is not
the vacuum state of the $a$ and $b$ oscillators but rather a two mode
squeezed vacuum given by
\begin{equation}
  \label{eq:GS}
  \ket{GS}=\frac{1}{\mathcal{N}}\prod_{k}\exp{\left(-\tanh\eta_k^+
      \alpha_{k}^{\dagger}\alpha_{-k}^{\dagger}-\tanh\eta_k^-
      \beta_{k}^{\dagger}\beta_{-k}^{\dagger}\right)}\ket 0
\end{equation}
where
$\alpha_k/\beta_k=\frac{1}{\sqrt{2}}\left(\pm
  e^{i\varphi\left(k\right)}a_{k}+b_{k}\right)$ are the
eigen-operators in the absence of the pairing terms and $\ket{0}$ is
the vacuum state of $a_n$ and $b_n$. In the weak coupling limit
$\eta_k^\pm\rightarrow 0$ and the ground state identifies with the
vacuum of the oscillators $\ket{GS}\rightarrow\ket{0}$. The zero point motion of the quantum oscillators is manifested in an
average mean photon number in the ground state. To lowest order in $v$
and $w$ it is given by
$\langle a^{\dagger}_n a_n \rangle = \langle b^{\dagger}_n b_n\rangle
= (v^2+w^2)/(2\epsilon_0^2)$.

The topology of one-dimensional systems with inversion symmetry is
characterized by the Zak phase~\cite{Zak1989}
\begin{equation}
  \varphi_{Zak}^\pm=i\int_{-\pi/l}^{\pi/l}{\bm{\psi}_k^{^\pm\dagger}
    * \partial_k \bm{\psi}^\pm_k dk}.
\end{equation}
Since the Bogoliubov transformation for boson is symplectic rather
than unitary as for fermions, the scalar product is modified $\bm{\psi}^\dagger *\bm{\psi}=\bm{\psi}^\dagger \mathcal{D} \bm{\psi} $.  For our choice
of basis (Eq. \ref{eq:basis}), the diagonal matrix $\mathcal{D}$ take the form
$\mathcal{D}=\text{diag}(1,-1,1,-1)$~\cite{xiao2009, Shindou2013_mag,
  xiao2009Bog}. The Zak phases of the two bands are identical and equal to 
\begin{equation}
  \varphi^\pm_{Zak}=\frac{1}{2}\int_{-\pi/l}^{\pi/l}{\partial_k
    \varphi(k) dk}
\end{equation}
independently of the Bogoliubov rotation $\eta_k^\pm$. Therefore the Zak
phase does not depend on the coupling strength. For $v>w$ the system
is in a trivial phase with $\varphi^\pm_{Zak}=0$ while for $v<w$ the
system is in a toplogical phase with $\varphi^\pm_{Zak}=\pi\ (\textrm{mod}\ 2\pi)$.  At the transition point $v=w$ the gap between the band closes.

\begin{figure}
  \begin{center}
    \includegraphics[width=0.49\textwidth]{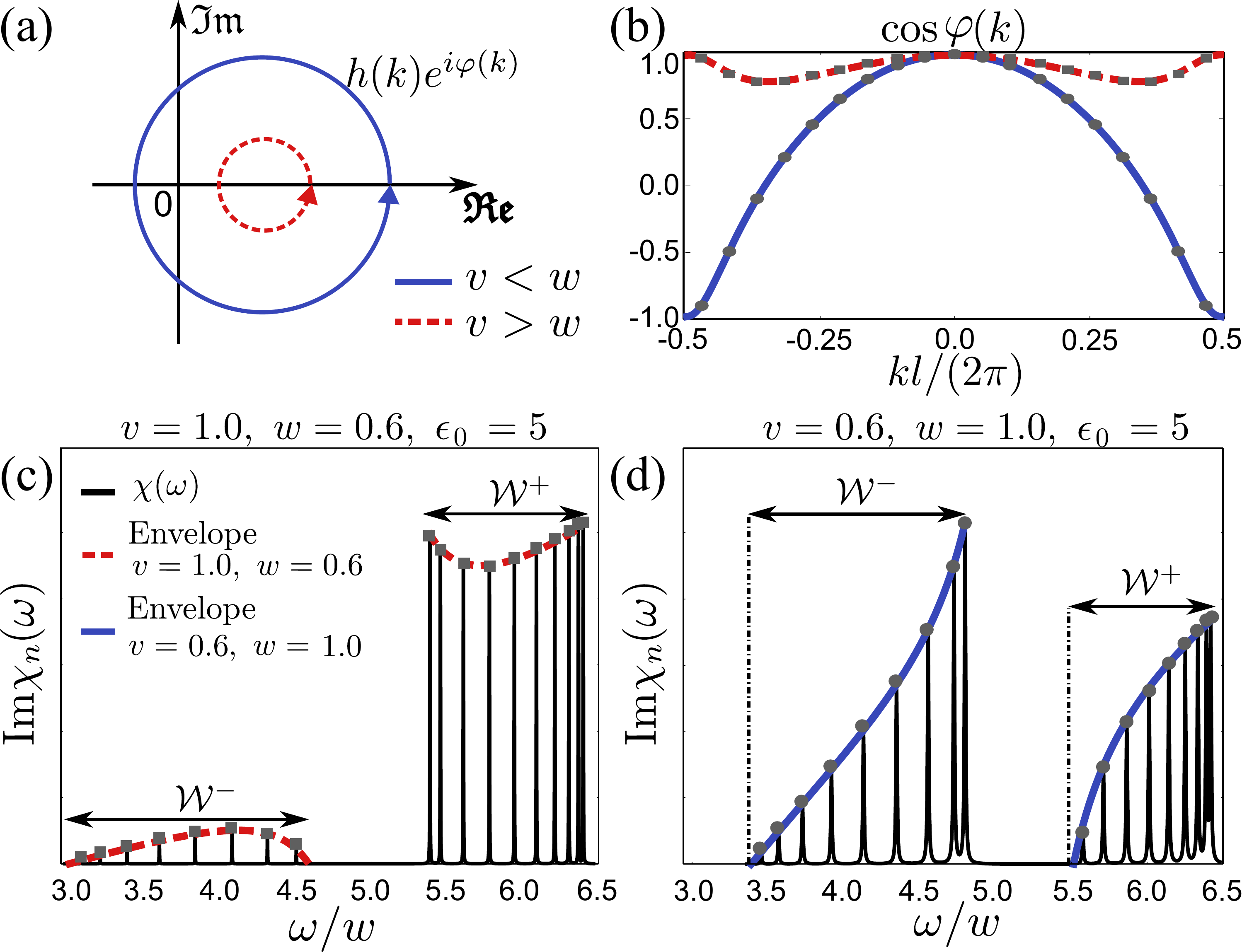}
  \end{center}
  \vskip -0.5cm \protect\caption[Topology measurement]{ (color online)
    \it a) The circle defined by $h(k)e^{i\varphi(k)}$ in the two
    phases: for $v < w$ it winds once around the origin, while for
    $v > w$ the winding number equals to zero. The topological phase
    transition occurs at $v = w$. b) $\cos\varphi(k)$ in the two
    phases. In the topological phase it is a surjective function onto
    [-1,1] while in the trivial phase it is limited to [0,1]. c-d) The
    correlation function $\chi_n(\omega)$ in Eq.~\eqref{eq:chi_n} in
    the trivial (c) and topological (d) phases for a ring composed of 20 unit cells. The envelope of the resonances reveals the winding of the Zak phase as explaned in the text.
    \label{fig:top_meas}
  }
\end{figure}

An equivalent and perhaps more transparent way to characterize the
topology of the system is by looking at the winding number of the
phase $\varphi(k)$, i.e. the number of times $e^{i\varphi(k)}$ winds
around the origin for $k\in [-\pi/l,\pi/l]$. The function $h(k)e^{i\varphi(k)}=v+we^{-ikl}$ defines a circle in the complex
plane of radius $w$ around $(v,0)$ (see Fig.~\ref{fig:top_meas}(a-b)). In
the topological phase $v<w$, the circle encloses the origin once and
$\varphi(k)$ is a surjective function onto $[-\pi,\pi)$. In the
trivial phase $v>w$, the circle does not enclose the origin and
$\vert\varphi(k)\vert<\pi/2$. Hereafter, we propose a method to observe the winding of the phase $\varphi(k)$
based on a reflection measurement.

The fact that the topology does not depend on the Bogoliubov rotation
$\eta_k^{\pm}$ can be understood using adiabatic switching of the pairing terms. A transition between two gapped topological phases implies closing of the gap at the transition point.  We consider a Hamiltonian which interpolates between the
SSH chain and our chain
\begin{align}
  H_{ad}
  =& \sum_n{\epsilon_0\left( a_n^\dagger a_n +b_n^\dagger b_n
            \right)} \nonumber
  \\
  +& \sum_n{\left(va_n b_n^\dagger+w a_{n+1}b_n^\dagger +h.c.\right)}
  \\
  +& \sum_n{\left( \delta v a_nb_n+\delta w a_{n+1}b_n +h.c.\right)
     }. \nonumber
\end{align}
For $\delta v=\delta w=0$ the Hamiltonian identifies with the SSH
chain Hamiltonian while for $\delta v=v$, $\delta w=w$ it yields
Eq.~\eqref{H_main}.  The spectrum of the Bogoliubov quasi-particles
for any $\delta v$, $\delta w$ is given by
\begin{equation}
  \epsilon_\pm^{ad}(k)=\sqrt{(\epsilon_0\pm h(k))^2-\delta h^2(k)}
\end{equation}
where $\delta h=\left\vert\delta v+\delta we^{-ikl}\right\vert$. For
any real $\delta v$, $\delta w$ the pairing terms do not close the
energy gap, therefore the topology remains the same as in the weak coupling regime.

Next we propose an experimental protocol to measure the Zak phase by probing the winding of $\varphi(k)$ which is based on the
measurement of the correlation function
$\chi_n(t)=-i\Theta(t)\langle [q(t),q(0)] \rangle$ of the single cell (charge)
operator $q_n=a_n^\dagger +a_n +b_n^\dagger +b_n$.
This correlation function can be probed by the reflection coefficient of
a transmission line (capacitively) coupled to a single cell in the chain
which is given by $r(\omega)=1+\lambda^2 \chi_n(\omega)$, 
here $\lambda$ is the (weak) coupling constant between the transmission line and the chain.
In the periodic boundary condition
$\chi_n(t)$ does not depend on $n$. The Fourier transform of the
correlation function yields
\begin{align}
  \label{eq:chi_n}
  \hspace{-15pt}
  \chi_n(\omega)
  =&
     \frac{1}{N}\sum_{k} \left(1+\cos\varphi\left(k\right)\right) \frac{\epsilon_0}{\epsilon_+(k)}
     \frac{1}{\omega-\epsilon_{+} \left(k\right)+i0^+}
  \nonumber \\
  +& \frac{1}{N}\sum_{k} \left(1-\cos\varphi\left(k\right)\right)
     \frac{\epsilon_0}{\epsilon_-(k)} \frac{1}{\omega-\epsilon_{-} \left(k\right)+i0^+},
\end{align}
where $N$ means the number of unit cells.

The imaginary part of $\chi_n(\omega)$ (See  Fig.~\ref{fig:top_meas}(c-d)) exhibits resonances at the eigen-energies $\epsilon_\pm(k)$, thus measuring the band structure of the chain. The strong-coupling limit can be probed through the asymmetry of the bandwidths (Eq.~\eqref{eq:bandwidths}). The envelope of the resonances reflects the winding of the phase $\varphi(k)$ through the terms $1\mp\cos\varphi(k)$. Let us consider the envelope function of the first band $1-\cos\varphi(k)$. The bottom of the first band is at $k=0$ where 
$\varphi(0)=0$ and its top is at $kl=\pi$ where $\varphi(\pi)=0$ in the trivial phase and $\varphi(\pi)=\pi$ in the topological phase. Therefore, in the topological phase the envelope function of the first band is a monotonically increasing function whereas in the trivial phase it starts and ends at zero. The behaviour of the envelope of the second band resonances can be understood in the same way, noticing that the bottom of the second band is at $kl = \pi$ and its top is at $k = 0$.
The general behaviour of the envelope function is similar in the weak and strong coupling regimes. 

A single $k$ mode can be observed by coupling the transmission line to
all the cells of the chain. The reflection coefficient is then related
to the correlation function $\chi_{k=0}(\omega)$ of
$q_{k=0}=\frac{1}{\sqrt{N}}\sum_n q_n$, given by the $k=0$ mode of
Eq.~\eqref{eq:chi_n}. The $k$ mode can be scanned across the Brillouin
zone in a way similar to Bloch oscillations by introducing an
effective gauge field \cite{roushan2016chiral}.

The bulk-edge correspondence asserts the existence of midgap localized states at the boundaries of a finite topological system.
In contrast to the bulk Bogoliubov excitations, these edge states
are particle like. We show that the SSH ansatz for the edge operators
\begin{align}
  \label{edges}
  \gamma_a
  &= \frac{1}{\mathcal{N}}\sum_n \left(-\frac{v}{w} \right)^n a_n
  \nonumber \\
  \gamma_b
  &= \frac{1}{\mathcal{N}}\sum_n \left(-\frac{v}{w}
            \right)^{N-n} b_n,
\end{align}
create eigenstates of the system with energy $\epsilon_0$.  For $N\gg1$, the commutation of the Hamiltonian and the
edge state operators is
$[H,\gamma_{a/b}^\dagger]=\epsilon_0\gamma_{a/b}^\dagger$. Therefore the energy of the edge states
$\ket{\psi_{a/b}}=\gamma^\dagger_{a/b}\ket{GS}$ is
\begin{equation}
  H\ket{\psi_{a/b}}=\epsilon_0\ket{\psi_{a/b}},
\end{equation}
where $\ket{GS}$ is the ground state of the system Eq.~\eqref{eq:GS}.
In the strong coupling regime, the edge-states are a many-body entangled state of the $a$ and $b$ operators.

\begin{figure}
  \begin{center}
    \includegraphics[width=0.49\textwidth]{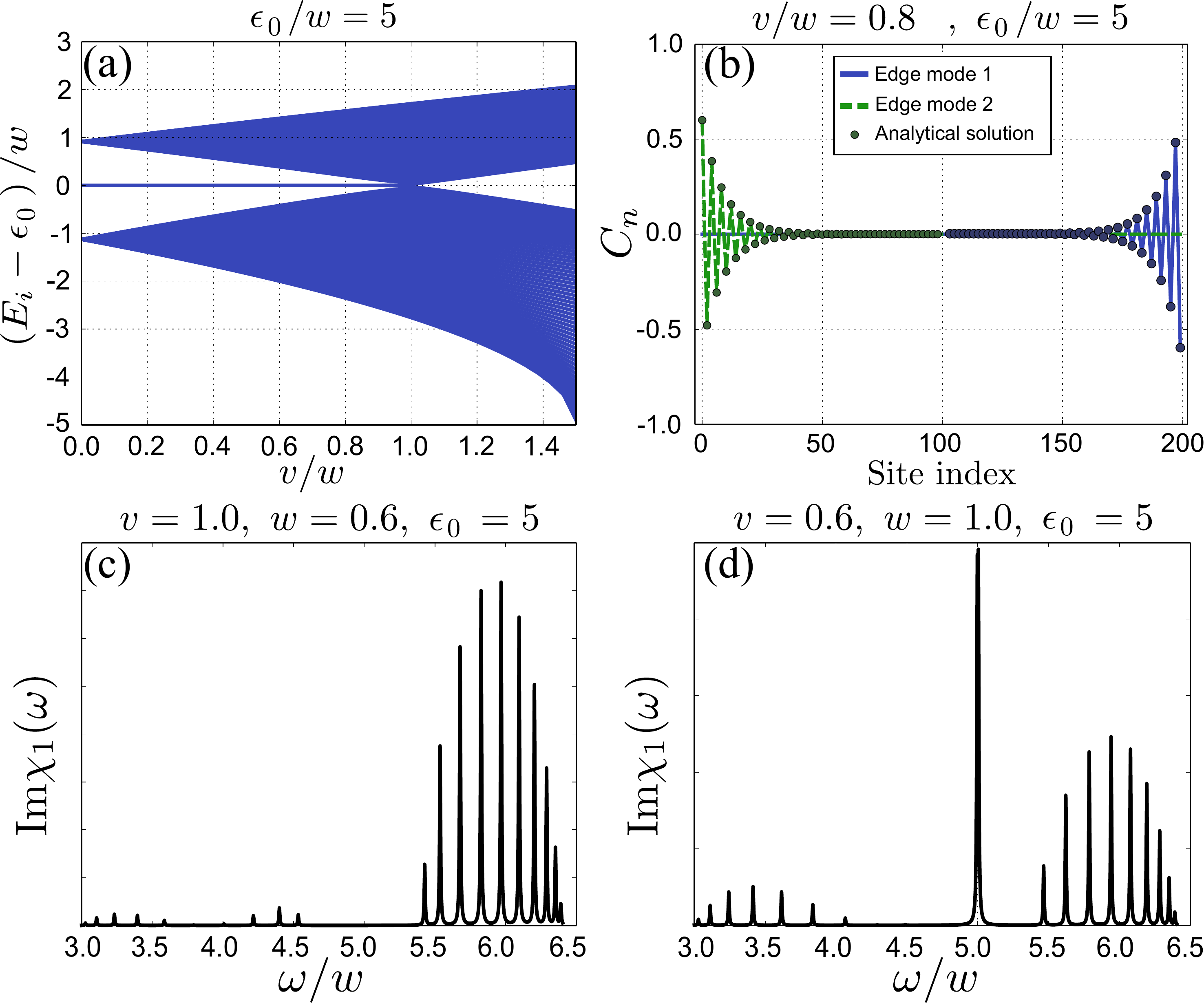}
  \end{center}
  \vskip -0.5cm \protect\caption[Edge states]{ (color online) \it a)  Energy spectrum of a finite chain with $N=100$ unit cells (200 sites) as a
    function of the hopping matrix elements $v/w$. Midgap energy
    states appear in the topological phase $v<w$.  b) Bogoliubov
    coefficients of the two midgap edge states and energy spectrum for the same chain.  c-d) The correlation function
    $\chi_{n=1}(\omega)$ in a system of 20 unit cells with open
    boundary conditions.
    \label{fig:edge_states}
  }
\end{figure}

It is worth noting that for different on-site energies $\epsilon_a$
and $\epsilon_b$ on two sublattices $A$ and $B$, the edge
states have the same form given by Eq.~\eqref{edges} but with their
corresponding energies $\epsilon_a, \epsilon_b$. Since addition of
such staggered on-site potential breaks inversion symmetry the Zak
phase is not quantized and a topological number cannot be defined in
that case. 

Fig.~\ref{fig:edge_states}(a) shows the appearance of the midgap states
in the energy spectrum of a finite system with $N=100$ unit cells in the
topological configuration. The spectrum was obtained by an exact
Bogoliubov diagonalization performed in real space
$\gamma_n=\sum_{n=1}^N\left( \mathcal{U}^a_na_n
  +\mathcal{V}_n^aa_n^\dagger+\mathcal{U}_n^bb_n+\mathcal{V}_n^bb_n^\dagger\right)$.
The Bogoliubov coefficients of the midgap states $\mathcal{U}_n^{a/b}$
are plotted in Fig.~\ref{fig:edge_states}(b), where the circles
correspond to our analytic solution for an infinite chain
Eq.~\eqref{edges}.

The midgap edge state can also be observed by the correlation $\chi_{n=1}$
for a system with open boundary condition. Fig.~\ref{fig:edge_states}(c-d) shows the imaginary part of $\chi_{n=1}$ for an open system with
$N=20$ unit cells. The midgap states appear only in the topological phase ($v<w$). Note that the envelope of the bulk resonances in $\chi_{n=1}$
differs from the periodic boundary condition case due to boundary
effects.

\begin{figure}
  \begin{center}
    \includegraphics[width=0.49\textwidth]{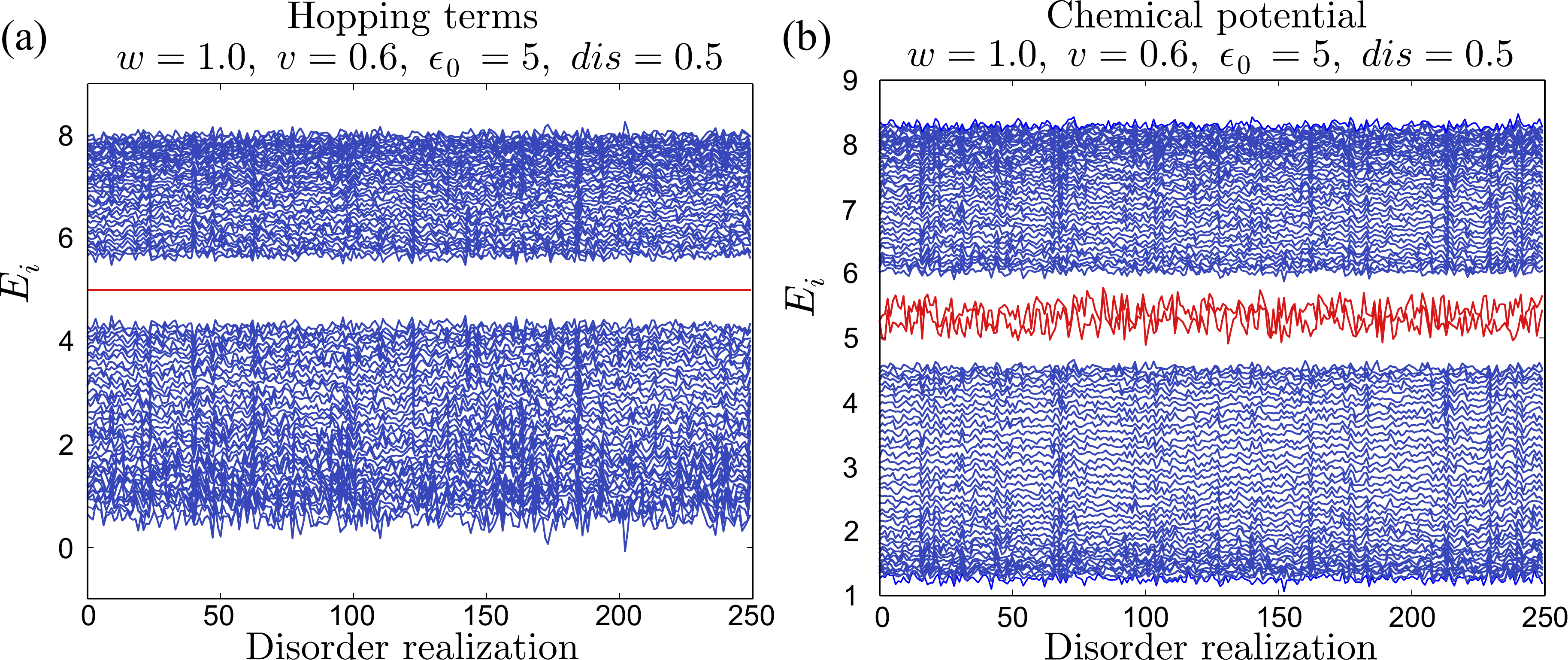}
  \end{center}
  \vskip -0.5cm \protect\caption[Energy spectrum for different
  disorder realizations]{ (color online) \it Band structure for
    different disorder realizations. a) Disorder in the hopping terms
    $v$, $w$. b) Disorder in the chemical potential $\epsilon_0$. The
    disorder is introduced by adding a noise term uniformly distributed
    between $[-dis,dis]$.
    \label{fig:dis}
  }
\end{figure}

Finally, we address the effect of disorder on the edge states. Disorder in the hopping terms $v$ and $w$ does not break the
sublattice symmetry, hence it does not change the midgap states energy and
modifies only the wave-function of the edge-states to
\begin{align} 
  \gamma_a &= \frac{1}{\mathcal{N}}\sum_n
             \left(\prod_{m=1}^n-\frac{v_m}{w_m}\right) a_n
  \\
  \gamma_b &= \frac{1}{\mathcal{N}}\sum_n
             \left(\prod_{m=N}^{N-n}-\frac{v_m}{w_m}\right)  b_n ,
             \nonumber
\end{align}
where $v_n$ and $w_n$ are the site dependent hopping coefficients. In
contrast, disorder in the chemical potential $\epsilon_0$ breaks the
sublattice symmetry and modifies the energies of the edge states. For
a large disorder these states can not be distinguished from the
bulk. Fig.~\ref{fig:dis} shows the eigen-energies of the system in the
topological phase for different realizations of disorder.

To summarize, we have shown that the topological nature of the dimer chain is preserved in the strong-coupling (quantum) regime and gives rise to the formation of topological Bogoliubov excitations as a result of the pairing term in the Hamiltonian. We have presented a
method to observe the Zak phase and the strong-coupling properties in the bulk. Our results are measurable with current technology \cite{Wurzburg,Guichard2015,Fitzpatrick,Takis}. As long as interactions produced by non-linearities (such as Bose-Hubbard models) do not break the sublattice symmetry, the topological properties would remain the same.

Acknowledgements: We acknowledge discussions with Camille Aron, J\'
erome Esteve, Julien Gabelli, Shyam Shankar, Lo\" ic Henriet, Lo\" ic Herviou,
Guillaume Roux and Ronny Thomale. This work has also benefitted from presentations and
discussions at the workshop on Quantum Sensing in Dresden and at the interdisciplinary worshop on topological phenomena in Lyon. This
work is funded by the DFG 2414 and by the Labex PALM.

\bibliographystyle{apsrev4-1}
\bibliography{Article_SSH_17nov_T.bbl}
\end{document}